\documentstyle[12pt,epsf]{ioplppt}
\begin{document}
\jl{4}
\title{An improved technique of image analysis for
IACTs and IACT Systems}[Improved image analysis for IACTs]

\author{M.~Ulrich, A.~Daum, G.~Hermann and W.~Hofmann}
\address{Max-Planck-Institut f\"ur Kernphysik, P.O. Box 103980,
D-69029 Heidelberg, Germany}

\begin{abstract}
In order to better utilize the information contained in
the shower images generated by imaging Cherenkov telescopes (IACTs)
equipped with cameras with small pixels, images are fit
to a parametrization of image shapes gained from 
Monte Carlo simulations, treating the shower direction,
impact point, and energy as free parameters. 
Monte Carlo studies for a 
system of IACTs predict an improvement of order
1.5 in the angular resolution.
The fitting technique can also be applied to 
single-telescope images; simulations indicate that
the shower direction in space can be reconstructed event-by-event
with a resolution of $0.16^\circ$ to $0.20^\circ$,
allowing to generate genuine source maps.
Data from Crab
observations with a single HEGRA telescope confirm this
prediction.
\end{abstract}

\maketitle

\section{Introduction}

Over the last decade, imaging atmospheric Cherenkov telescopes have 
proven the prime instrument for $\gamma$-ray astronomy in the TeV 
domain~\cite{review}. 
Both the orientation and the shape of Cherenkov images are
exploited to supress cosmic-ray background in the search for point
sources of $\gamma$ rays: 
$\gamma$ shower images seen in the camera point
back to the source location, and are characterized by narrow, compact
images. In contrast, cosmic rays generate hadronic showers with 
wider and more diffuse images, and random orientation. 

\begin{figure}[hb]
\begin{center}
\mbox{
\epsfysize7.5cm
\epsffile{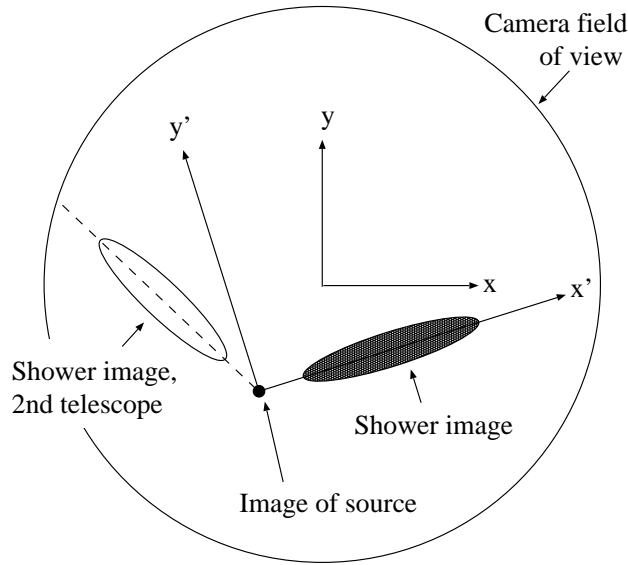}}
\end{center}
\caption
{Image of a shower in the camera, with the coordinate system
$(x,y)$ fixed to the camera, and the system $(x',y')$ defined
by the source image
and the direction to the shower impact point 
(after accounting for the reversal of signs occuring for the
mirror image).
Also illustrated is the reconstruction of the
shower direction using images from multiple telescopes.}
\label{fig_geometry}
\end{figure}

Recent
improvements of the imaging Cherenkov technique include the
use ``high-resolution'' cameras with pixel sizes of $0.15^\circ$
or less, capable of resolving fine details of the image,
and the stereoscopic technique, where a shower is observed
simultaneously by several telescopes, allowing to geometrically
reconstruct the shower axis, and hence the direction of the 
primary particle. Since Cherenkov shower images in the camera 
point to the 
image of the source, the apparent source of air shower
can be reconstructed by superimposing the images of several
cameras and intersecting the image axes (Fig.~\ref{fig_geometry})
\cite{kohnle_paper,crab_stereo,whipple_source}.
Opposite to the source image, the shower image
points towards the location where the shower axis intersects
the plane of the telescope dish. The impact location can be
therefore be derived by extrapolating the image axes,
starting from the locations of the telescopes, until the
lines intersect at one 
point
\footnote{This simple method
applies only if all telescope dishes lie in a plane 
perpendicular to the telescope axis; the extension to
the general case is however straightforward.}.

Cherenkov images are
traditionally described by parameters related to the first and
second moments of the intensity distribution in the 
camera~\cite{hillas_param}: 
the center of gravity, the length of the major and minor axis
of the image ``tensor of inertia'', and its orientation. While rather
powerful, this method has two shortcomings: it does not make use of the
full information obtained with todays 
cameras, where a typical shower lights
up twenty and more  pixels, and it requires the use of a ``tail cut''
to eliminate pixels which do not belong to the image, yet show
some signal due to night-sky background light. Since this tail cut
is usually set at a level around 5 photoelectrons, tails of the image
are excluded, too. 

To provide tools for an improved image analysis,
we have, over the last years, developed a 
technique~\cite{ulrich,ulrichphd}
 which makes better
use of the information contained in the image, by fitting the 
observed intensities to a model of Cherenkov images, with the showers
characteristics - direction, core location, and energy - as free 
parameters. In this paper, we describe first the simulation and 
parametrization of the images, then the fitting procedure and its
predicted performance when used in a system of imaging Cherenkov
telescopes.
Finally, we apply the fitting technique also to single-telescope data
and demonstrate its performance 
using images
obtained with one of the HEGRA Cherenkov telescopes during observations
of the Crab Nebula. Emphasis on the present work is on this
interpretation of single-telescope images; multi-telescope
data has become available, but the analysis is still in the
early stages~\cite{crab_stereo}.

A similar reconstruction
technique has been studied by the CAT group~\cite{cat_reco}.

\section{Modeling of Cherenkov images}

To perform a fit of the intensity profile of Cherenkov images,
an analytical parametrization of images was derived on the basis of
Monte Carlo simulations. The relevant parameters of the shower are its
impact distance $R$ from the telescope, the shower energy $E$, the
zenith angle $\Theta$ of the shower, and, to a certain 
extent, the height $h_{max}$
of the shower maximum. Other shower parameters, such
as the orientation of the shower axis, or the direction towards the location
of the shower core, correspond simply to translations and rotations of
the image; images are hence described in a coordinate system $(x',y')$
with its origin at the image of the source, and the $x'$-axis pointing
towards the shower impact point in the plane defined by the telescope dish
(see Fig.~\ref{fig_geometry})
\footnote{All camera coordinates are converted to photon slopes,
or equivalently, angles. For a given position $(X,Y)$ of a photon,
$(x,y) = (X/f,Y/f) \approx (\theta_x, \theta_y)$, with $f$ denoting 
the focal
length.}. Neglecting for the moment the
$h_{max}$-dependence, the photoelectron density in the image is 
given by a function $\rho(x',y'|R,E,\Theta)$, which in the following will
be factored into a longitudinal profile and a transverse profile,
$\rho(x',y') = \rho_{L}(x') \rho_{T}(y'|x')$, with
$\int \rho_{T}(y'|x') dy' \equiv 1$.

The parametrization refers to a camera with infinite granularity; to
compare with an actual image, the parametrized images are shifted and
rotated according to the presumed shower orientation, and the image 
intensity is integrated over the area of each of the camera pixels.
Since image shapes vary slowly with shower energy, the simulation effort
was simplified by generating events at one typical energy $E_0$, 
and scaling the intensities afterwards
according to energy. Similarly, the dependence on the zenith angle 
was introduced in a simplified fashion: only vertical showers 
($\Theta = \Theta_0 = 0$) were
considered, and image sizes (for fixed $R$) were later scaled with
$\cos \Theta$, reflecting the increasing distance of showers from the
camera for larger $\Theta$. 
Since most analyses so far concentrated on small zenith angles
(less than $30^\circ$), this assumption works sufficiently well.

\subsection{Simulation of Cherenkov images}

The simulation code~\cite{plya_mc,kono_mc}
 tracks electromagnetic and
hadron-induced cascades and includes the propagation of Cherenkov light
in the atmosphere, and the wavelength-dependent characteristics of 
the telescopes, such as reflectivity of the mirrors, the efficiency of
the light collectors in front of the PMTs, and the PMT quantum efficiency.
Image intensities and image characteristics have been compared with
other simulation codes, and good agreement has been found~\cite{system1}. 
The simulations use the characteristics of the HEGRA
telescopes~\cite{hegra_system,hermann_padua}, 
in particular the mirror area of 8.5 m$^2$ and the location at
of 2200~m a.s.l. To parametrize the images, 10000 vertical showers 
were generated
and observed with telescopes positioned at 15 distances between 0 and 300~m
from the shower axis. In case of $\gamma$ ray initiated showers, a fixed
energy $E_0 = 1$~TeV was chosen. For studies concerning hadron 
rejection, also
proton showers were generated; here, $E_0 = 3$~TeV was used, 
resulting in roughly
the same number of photoelectrons in a typical image.

\subsection{Longitudinal image profile}

\begin{figure}[htb]
\begin{center}
\mbox{
\epsfysize9cm
\epsffile{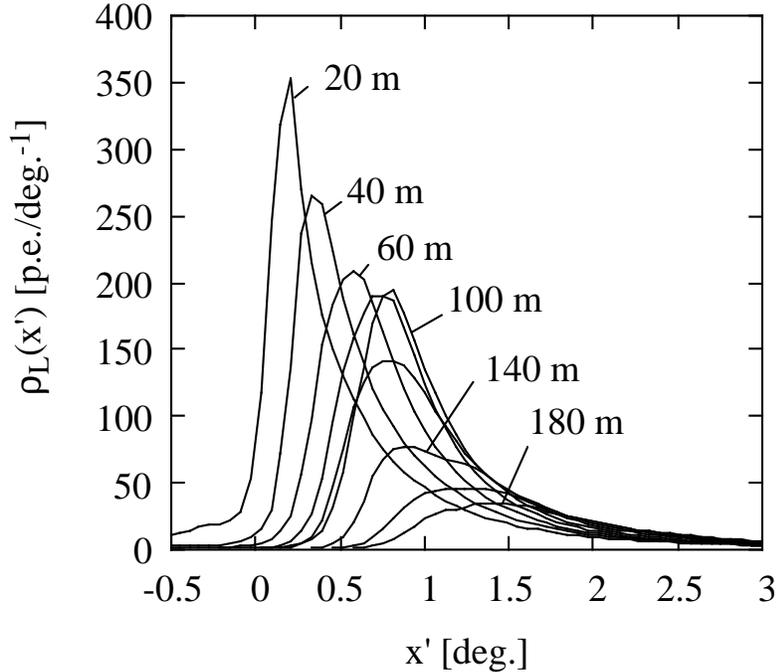}}
\end{center}
\caption
{Longitudinal image profile for 1 TeV $\gamma$-ray showers at different 
distances $R$ = 20,40,60,80,100,120,140,160,180 m
between shower axis and telescope, as determined from 
Monte-Carlo simulations.}
\label{fig_dndx}
\end{figure}

The longitudinal image profile $\rho_{L}(x'|R,E_0,\Theta_0)
= \rho_{L}(x'|R)$ 
is obtained by projecting the image onto
the $x'$ axis, and averaging over many showers.
We note that the coordinate $x'$
measures essentially the height of emission $h$ of a photon, $x' \approx R/h$,
up to effects caused by the finite radial extent of the shower. 
Comparing the profiles of individual showers with the average profile
(for a given $R$), one notices that the average profile does not 
describe the individual images very well; they tend to be narrower.
The explanation is simply that the height of the shower maximum, and
hence the location of the peak of the longitudinal profile fluctuates
from shower to shower. The typical rms variation of one
radiation length in the
height $h_{max}$ of the shower maximum translates 
into a variation
in $x'$ of about 15\% (for a typical height of the shower maximum 
of 6~km above ground). Since a parametrization is needed which
applies to individual rather than average images, the height of 
the shower maximum of each simulated shower was determined from the 
evolution of the number of shower particles with depth, and all
images were scaled to the same average height of the shower maximum,
before averaging over the individual longitudinal profiles. In the
fits of Cherenkov images for multi-telescope observations, the height
$h_{max}$ was correspondingly included as a free parameter, which 
manifests itself as a scale factor for the image sizes. 
Since for
stereoscopic shower observations the impact point is strongly
constrained by the geometrical relations discussed earlier, both
the impact point and the height of the shower maximum can be 
determined separately. Fig.~\ref{fig_dndx} shows the resulting
longitudinal profiles for different impact distances.
We note that
this type of parametrization has one disadvantage: in some distance
ranges, a peak or shoulder in the angular distribution of photons
at the typical Cherenkov angle reflects emission in the early stages of the 
shower, where the smearing of angular distributions due to multiple
scattering of shower particles is small; due to the rescaling of angles,
this structure is smeared out. 

To describe the longitudinal profile as a function of $x'$
and $R$, a two-dimensional spline function was fit
\cite{nag} to the simulation
data, with the number and location of node points (19x17) optimized to
obtain a sufficiently good representation.

\subsection{Transverse image profile}

\begin{figure}[htb]
\begin{center}
\mbox{
\epsfysize10cm
\epsffile{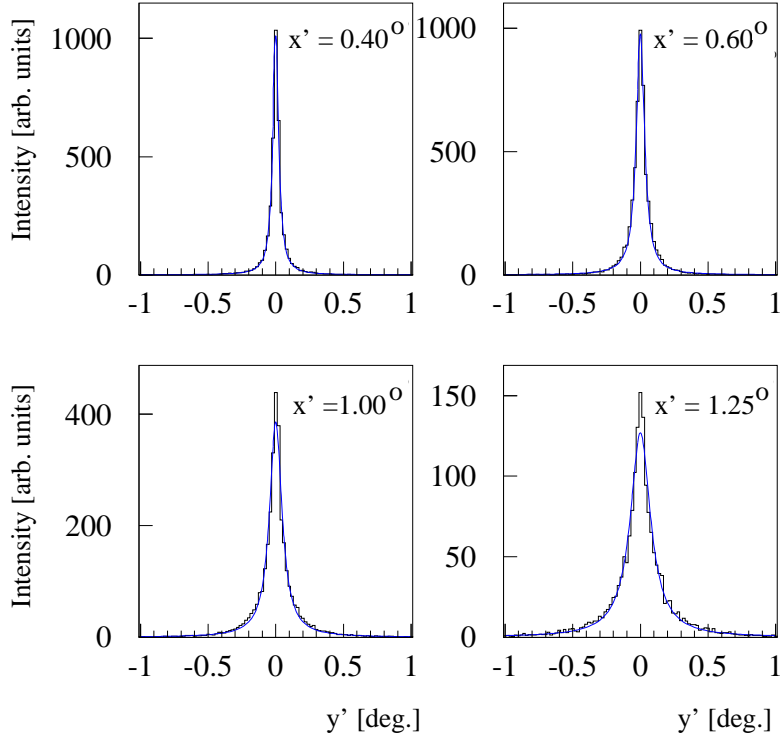}}
\end{center}
\caption
{Transverse image profiles for $R = 60$~m and different values of
$x'$, as obtained from Monte-Carlo simulations,
together with the fit function.}
\label{fig_dndy}
\end{figure}

\begin{figure}[htb]
\begin{center}
\mbox{
\epsfysize9cm
\epsffile{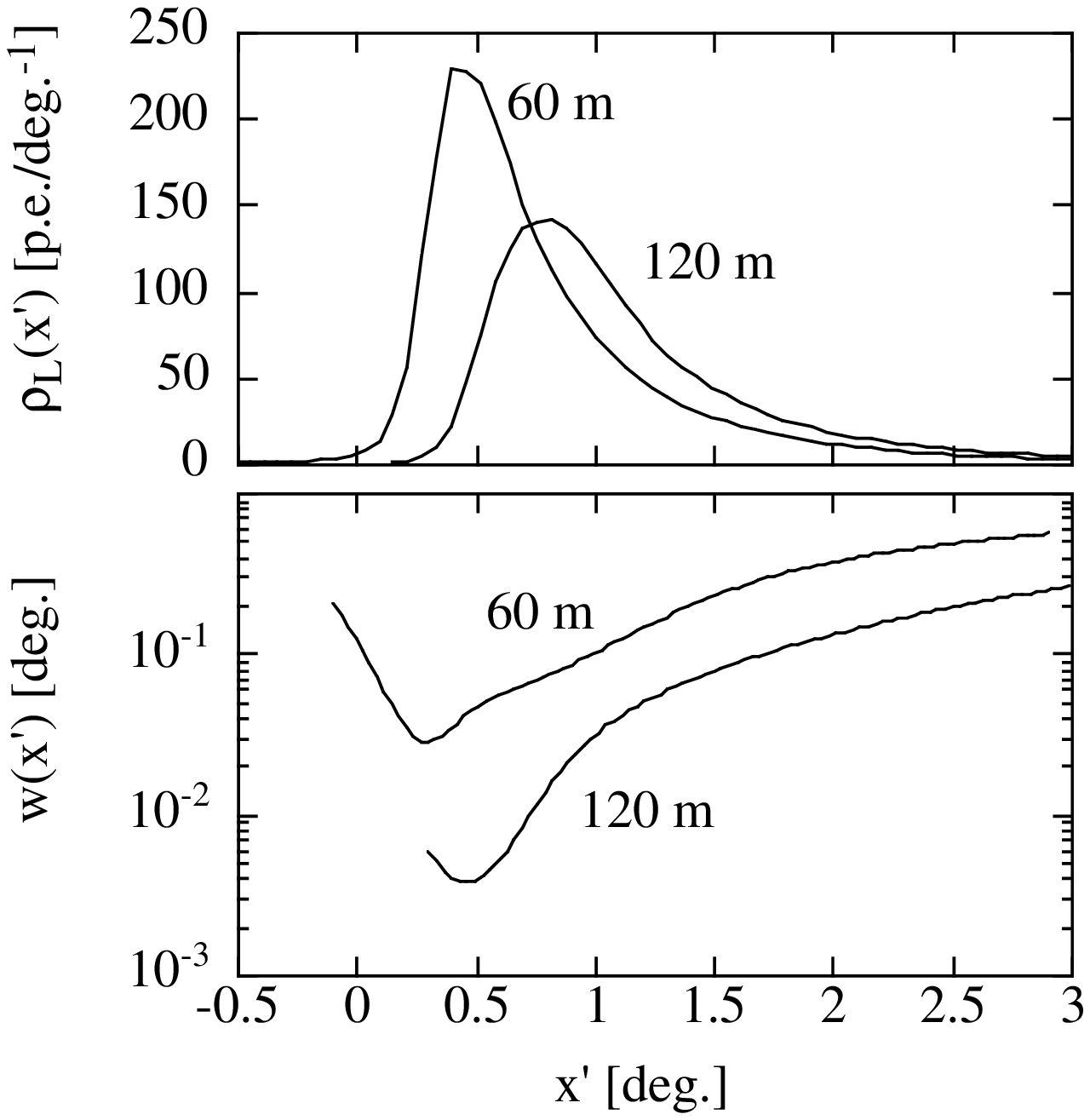}}
\end{center}
\caption
{Width parameter $w(x',R)$ of the transverse distribution as a function
of the longitudinal coordinate $x'$, for core distances $R$ of 60~m
and 120~m. For reference, the corresponding longitudinal 
distributions $\rho_L(x'|R)$ are also shown.}
\label{fig_fits}
\end{figure}

The transverse width of the Cherenkov image largely reflects the width
of the shower disk, or equivalently the angular distribution of shower
particles. One therefore sees a very narrow image at small values of 
$x'$, corresponding to large heights, and a wide image at large values
of $x'$, corresponding to the tail of the shower. Samples of transverse
profiles are shown in Fig.~\ref{fig_dndy}.
For all distances $R$ and relevant values $x'$, the transverse profile
$\rho_{T}(y'|x',R)$ is well described by the 
one-parameter function
$$
\rho_{T}(y'|x',R) = {1 \over \pi}
{w(x',R) \over y'^2 + w^2(x',R)}
$$
where the width of the profile is characterized by $w$. Fig.~\ref{fig_fits}
illustrates how $w$ varies with $x'$, for two different shower distances
$R$. We note the pronounced asymmetry in the image with respect to
its center of gravity along $x'$, both in the 
longitudinal profile, and in the width of the transverse profile. This
asymmetry is ignored in the conventional 
second-moment image parameters.
An interesting observation is that the shape of the transverse image 
profile illustrates why tail cuts are absolutely necessary in the second
moment image analysis, even in the absence of night-sky background:
with a profile varying like $1/y'^2$ for large $y'$, the second moment 
does not converge without some kind of cutoff, either in amplitude 
or directly in $y'$!

For use in the image fits, the width $w(x',R)$ is again described by a 
two-dimensional spline function.

\subsection{Fluctuations of image intensitities}

In order to properly fit Cherenkov images and to obtain error estimates
for the shower parameters, it is not sufficient to provide a parametrization
of image profiles as a function of the relevant parameters, one needs
also to know the fluctuations of the intensities measured in the 
individual pixels of the camera. With the fit based on a $\chi^2$
minimization, strictly speaking one needs the covariance matrix
of the amplitudes in all pixels. In particular, longitudinal
profiles of electromagnetic showers in matter are known to exhibit large 
correlated fluctuations; 
the energy deposition in nearby sections shows
a positive correlation, whereas the energy deposition in early and late
slices of showers is anticorrelated. Most of this correlation, however,
can be traced simply to the fluctuation in the location of the shower
maximum; once this effect is removed, correlations are modest. The 
dependence of longitudinal profiles on the shower maximum is however 
already explicitly taken into account in our model, 
by treating it as a free
parameter in the fit. We therefore considered the
pixel amplitudes as independent, and included in the errors
a term describing the Poisson fluctuation in the number, a term
accounting for PMT gain fluctations (using a width of the
single-photoelectron peak of about 70\%), and a term accounting
for night-sky background and electronics noise (about one photoelectron
in case of the HEGRA telescopes). 

\section{Fitting of Cherenkov images}

The fitting technique to derive shower parameters was initially
developped to be used for system of Cherenkov telescopes; more 
recently, it was successfully applied to images from single telescopes.
In this chapter, we briefly summarize the technique and the simulation
studies carried out for the HEGRA system of Cherenkov telescopes.

\subsection{Fit procedure}

Shower parameters are optimized to best describe the measured images
by numerically minimizing the $\chi^2$ function
$$
\chi^2 = \sum_{j} \sum_{i} 
w_{i,j} {\left( a_{i,j} - (E/E_0) \int_{\mbox{pixel}} 
\rho(x',y') dx'dy' \right)}^2~~~.
$$
Here, $a_{i,j}$ is the amplitude measure in the $i$-th pixel of the 
$j$-th camera, and the weight $w_{i,j}$ is calculated based on the
expected fluctuation of pixel amplitudes, $w_{i,j} = 1/\sigma^2_{i,j}$.
The predicted amplitudes are obtained by integrating the intensity
profile over the pixel area. The minimization and the error estimates
for the fit parameters are
carried out with standard routines~\cite{nag}, treating the
shower direction (2 parameters), the impact point (2 parameters), the
energy, and the height of the shower maximum as free parameters.

In order to achieve rapid convergence of the fit, the choice of good 
starting values is essential. For the shower direction and impact
point, the starting values were derived using a conventional geometrical
reconstruction. In cases where only two telescopes contribute,
direction and impact point are directly given by the intersections
of the image center lines; for events with more telescopes,
the shower parameters were determined separately from each pair
of telescopes, and averaged, weighted with the sine of the angle 
between the views.
Given these starting values, and the assumed linear
dependence between light yield and energy, a starting value for the 
energy $E$ can be derived analytically. The average height of the
shower maximum is used to seed the height of the shower maximum.

Since the integration over pixel area is partly numerical and 
time-consuming, the number
of pixels included in the determination of $\chi^2$ was limited to
those containing potentially relevant information, plus a boundary
region. These boundary pixels should contain no signal, but are
required to ensure the stability of the fit
\footnote{It may be worth to note that, unlike for the conventional
tensor analysis, it does not hurt the quality of the fit if additional
pixels beyond the shower image are included with their proper weights. 
As long as no signal
is predicted for a pixel, this pixel simply adds a constant term
to the $\chi2$, and does not change the location of the minimum or
the errors associated with the parameters.}.
Typically, all pixels were included which were within an
ellipse with half-axes equal to 5 times the {\em width} and {\em length}
Hillas image parameters; for events with very small width or length,
a fixed region was used.

\subsection{Fitting of multi-telescope images}

The simulation studies were carried out for a system similar to
the HEGRA CT system \cite{hegra_system,hermann_padua}
under installation
on the Canary Island of La Palma, 
at the Observatorio del Roque de los Muchachos
of the Instituto Astrofisico de Canarias.
The telescopes are located at 2200~m a.s.l., have 8.5~m$^2$ mirror area and 
are equipped with 271-pixel cameras, with a pixel size of $0.25^\circ$
and a hexagonal field of view of $4.3^\circ$ effective diameter. 
Four telescopes form the corners of a square with about 100~m sides,
the fifth telescope is located in the center of the square. The 
simulations used PMT quantum efficiencies measured for an earlier 
generation of PMTs; the actual PMTs used are about 30\% more
efficient and hence thresholds should be scaled down correspondingly.
The simulation implemented a trigger condition close to one used in
HEGRA: at least two pixels of a camera need to detect 10 or more
photoelectrons to fire the telescope trigger, and at least two telescopes
need to trigger to accept the event. Two modes of analysis were
considered. In one mode, only telescopes which had triggered were
included in the analysis; in the other mode, also telescopes which did
not trigger, but had images with at least 20 photoelectrons were included 
(the
HEGRA telescopes with their Flash-ADC readout system and signal storage
allow to recover the signal even if a telescope did not trigger). 
In the following, resolutions for the shower direction and the
impact position are given for both modes. The resolutions quoted
correspond to the diameter of a circle containing 68\% of the 
events.

\begin{figure}[htb]
\begin{center}
\mbox{
\epsfysize12cm
\epsffile{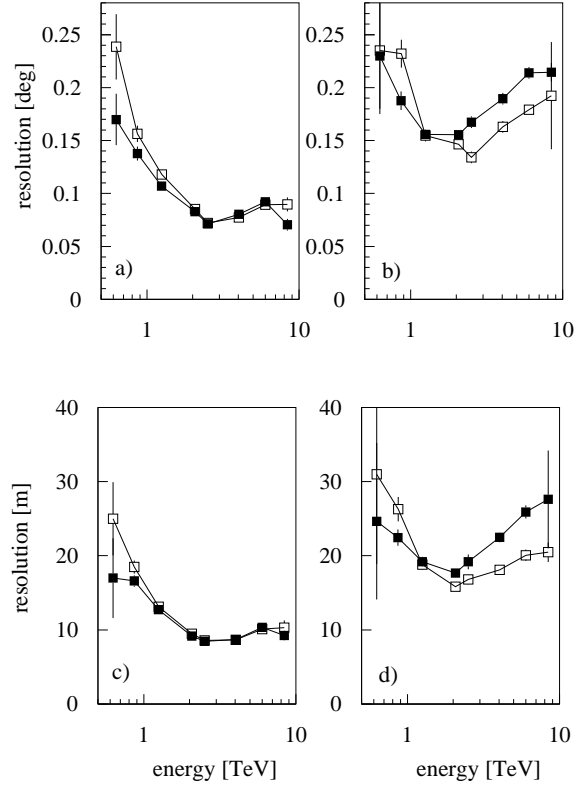}}
\end{center}
\caption
{Resolution 
in the reconstruction of the shower direction
(a),(b) and of the shower impact position (c),(d) using a 
five-telescope system,
shown as a function of energy. The resolution is defined such that
68\% of all events are contained within a circle of the given
diameter. The resolutions are shown both using the fitting
technique (a),(c), and the conventional image analysis (b),(d).
Open symbols: only triggered telescopes included in the 
analysis (2 pixels above 10 photoelectrons). Full symbols:
also including non-triggered telescopes with at least 
20 photoelectrons in the image.}
\label{fig_sys}
\end{figure}

Fig.~\ref{fig_sys} displays the angular resolution of the telescope
system as a function of energy, both for fitting technique (a)
and the conventional geometrical reconstruction (b). One notices
a significant improvement if the image fit is used. In particular at small 
energies, adding non-triggered telescopes improves the fit further.
A surprising feature is that the angular resolution worsens for large
energies, particularly dramatic in case of the conventional reconstruction.
The reason is that showers at increasingly larger distances contribute,
and hence the stereo angles under which the telescopes view the shower
decrease, correspondingly increasing the reconstruction errors. If
only showers with impact points within 120~m from the center of the
system are included, the angular resolution improves to better than 
$0.05^\circ$ at 10 TeV in case of the fit. Fig.~\ref{fig_sys}(c),(d)
shows the resolution in the reconstruction of the shower impact point;
again, the fit provides a clear improvement over the simple
geometrical reconstruction. A good localization of 
the impact point is important to allow reliable energy estimates.
The height of the shower maximum is reconstructed with an rms error
of 0.9 radiation lengths; the energy resolution is typically 20\% to 
25\%,
almost independent of the energy of the incident photon.

Overall, the fitting technique provides an improvement in resolution 
by typically a factor 1.5, and up to 2 in special regimes.

\subsection{$\gamma$-hadron separation using the image fit}

We had initially hoped that the fitting technique would also
provide significant improvements in the $\gamma$-hadron separation
based on image shapes (as opposed to image orientation). Various
techniques were tried, such as cuts on the quality ($\chi^2$)
of the fit with the $\gamma$-shower parametrization, or the fit of
images with both a $\gamma$-shower model and an analoguous proton
shower model, and cuts applied to the absolute $\chi^2$ as well
as to the $\chi^2$ difference between the two fits. To judge the
quality of a scheme, the enhancement in the signal-to-noise ratio
was used, $\epsilon_{\gamma}/\sqrt{\epsilon_{CR}}$. While the various
techniques did improve the signal-to-noise of $\gamma$-signals,
none of the variants was clearly superior to the classical image
cuts, which typically select $\gamma$-candidates as events with small
{\it width} and {\it length}, a large {\it concentration} of the signal
in a few pixels, and possibly a small amount of clutter outside the
main image cluster. It appears that the remaining correlations between pixels
and the non-gaussian signal fluctuations hurt the fit in this respect,
and are more effectively captured by the classical global event variables.
Since the classical variables are much faster to calculate, 
we have in most applications and studies pre-selected the 
$\gamma$-candidates 
based on the normal shape parameters, and then applied the
fit to best reconstruct the shower parameters, i.e. the
direction, impact point, and energy.

\section{Shower reconstruction for single-telescope images}

For Cherenkov observations with a single imaging telescope, the 
location of the source, i.e. the direction of the primary, is
constraint to lie on the image axis (see Fig.~\ref{fig_geometry}).
However, its position along this axis is not fixed. 
For this reason, Cherenkov telescopes do usually not provide 
images corresponding to those of optical telescopes, where a
source shows up as a peak in a two-dimensional map. Rather,
a location of a source is assumed, and it is checked if there
is an excess of events consistent with this source location.

To remedy this situation, it is tempting to apply the image fit to 
single-telescope images. The shape (length) of the image does
contain information about the impact distance $R$, and hence
about the distance between the peak of the Cherenkov image and 
the image of the source (see Fig.~\ref{fig_dndx}). 
The pronounced asymmetry of the image
along $x'$ allows to distinguish on which side of the image
the source is likely to be located. The image fit should be able
to extract this information in an optimum fashion. In addition,
it provides an improved reconstruction of the image axis and accounts
for image distortions caused by the outer edge of the camera,
which tend to severely distort conventional image parameters once images 
approach the camera border. 

Applied to a single image, the fit
cannot distinguish well between the influence of the impact distance
$R$, and variations in the height of the shower maximum $h_{max}$,
since to first order the image depends only on $h_{max}/R$. Therefore, in our
single-telescope fits, the height of the shower maximum was frozen.
As starting values for the shower parameters, a source location
separated from the center of gravity of the image 
by 3 times the image {\em length} along the image axis was used.
There is a two-fold ambiguity in the choice of the starting 
value. We usually chose the point closer to the center of the camera,
but also tried a variant where the fit was performed for both
starting values, and the best fit was chosen as the final result.

Monte-Carlo studies were again carried out for telescopes
of the HEGRA type, this time with a trigger condition 
requiring two pixels above 15 photoelectrons, as required
for single-telescope running in order to limit noise triggers.
The simulation showed that IACT images contain indeed 
enough information to estimate the impact parameter of the 
shower and hence the distance between the center of the shower image,
and the image of the source, with useful accuracy. 
In the $x'$ direction - along the image axis - the fit reconstructs 
the shower direction with a resolution
between $0.13^\circ$ and $0.17^\circ$,
depending somewhat on the criteria for event selection. 
The resolutions quoted are obtained from a gaussian fit
to the $x'$-distribution of reconstructed shower axes.
The resolution
in the transverse ($y'$) direction is significantly better,
around $0.08^\circ$ to $0.09^\circ$. Alternatively, one 
can quote a resolution in space, defined as the radius of
a circle containing 68\% of the reconstructed events. The
resulting values range between $0.17^\circ$ and $0.23^\circ$
and are slightly larger than expected based on the Gaussian
widths in $x'$ and $y'$, indicating slight non-Gaussian tails in
the resolution function. The fit provides an average 
energy resolution of 22\%.

At first sight, it may seem surprising that the resolution in
$x'$ is only a factor two worse than the resolution in $y'$ 
(which corresponds to the width of the distribution of the
conventional {\em miss} parameter). However, it does not even
require the admittedly not very transparent fitting procedure
to extract the information on the distance $d$ between the center
of the image, and the image of the source. Among the 
conventional image parameters, the ratio of {\em width} over
{\em length} of the image shows an almost linear depedence on $d$
and can be used to estimate $d$ with an rms precision of about
$0.25^\circ$ (see also, e.g., \cite{whipple_wl}). 
The use of the ratio of the {\em width} and 
{\em length} parameters has the advantage that the strong dependence
of the individual parameters on the image amplitude cancels to a 
large extent. Third moments of the longitudinal image profile
can be used to distiguish the `head' and `tail' of the shower with
reasonable reliability, i.e. allow to resolve the two-fold ambiguity 
in the source location. In this sense, the fitting
procedure actually provides more of a quantitative rather than a 
qualitative improvement of the image analysis. Nevertheless,
the extraction of the shower direction from single-telescope
images represents a significant step, and it was felt that
such a claim should not be supported by simulation studies
alone. The analysis was therefore applied to data gained in
observations of the Crab Nebula with the HEGRA telescope CT3,
the first of the HEGRA
telescopes to be equipped with a 271-pixel camera
\cite{hermann_padua}.

\subsection{Application to single-telescope Crab data}

The Crab data set used here comprises  80~h of observations, equally
distributed between on-source and off-source runs, resulting in
a total of $1.2 \cdot 10^6$ events at zenith angles below 
$45^\circ$. This data set shows a clear signal for TeV
$\gamma$-ray emission (\cite{crab_stereo}, Fig. 1).
 
\begin{figure}[htb]
\begin{center}
\mbox{
\epsfysize14cm
\epsffile{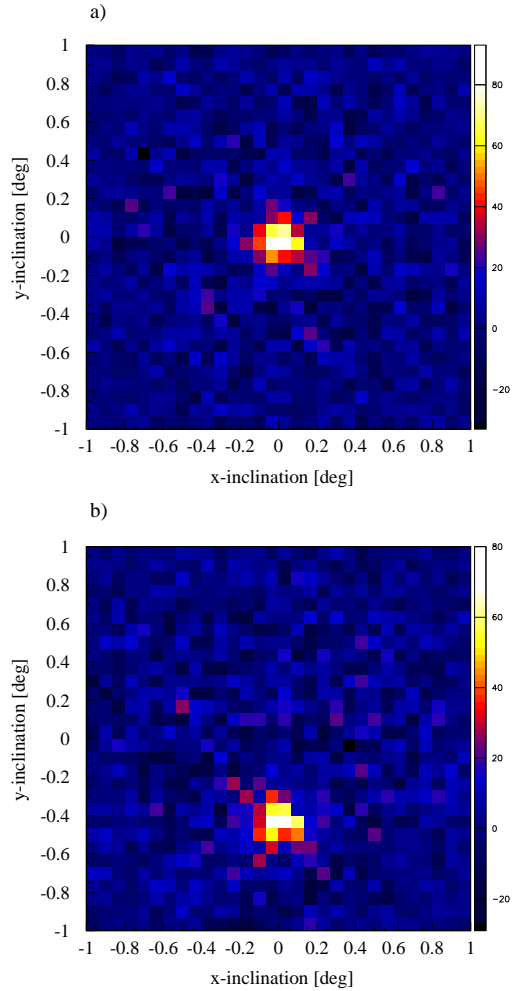}}
\end{center}
\caption
{(a) Distribution of the direction of fitted $\gamma$-ray
showers relative to the direction to the Crab Nebula
(located at the center of the camera), after 
subtraction of the off-source sample. (b) Same sample, but reconstructed
using a ``shifted'' camera, in which the Crab Nebula is
$0.42^\circ$ off-center.}
\label{fig_map}
\end{figure}

Because of an initial lack of readout channels of some of the pixels near
the fringe of the camera, analyses of this data set were mostly
restricted to the central 169 pixels. Events were collected
with a trigger condition requiring two pixels with signals above 15
photoelectrons. 
In a first analysis step, events were
preselected on the basis of the conventional second-moment
image parameters. The cuts included a {\em distance} cut at 
$1.1^\circ$ or $1.3^\circ$
to exclude heavily truncated images, and the requirements
{\em width} $<~0.13^\circ$, $0.15^\circ~<$ {\em length} $<~0.35^\circ$,
and {\em concentration} $>~0.4$. The selection reduced the
sample by a factor 6.
In the next step,
the images were fit. No further cuts on the fit quality were applied,
since the preselected images showed acceptable values
of $\chi^2$ of the fit, and on the basis of Monte-Carlo studies no further
improvement in $\gamma$-ray selection  was expected. Fig.~\ref{fig_map}(a)
shows the distribution of the reconstructed shower directions for
the Crab data sample,
after subtraction of the off-source data set. A clear excess at the 
center of the camera, at the location of the Crab Nebula, is evident,
with a characteristic width of about $0.2^\circ$. This `true'
source image from one of the HEGRA Cherenkov telescopes nicely 
demonstrates the power of the fitting technique.

\begin{figure}[htb]
\begin{center}
\mbox{
\epsfysize6cm
\epsffile{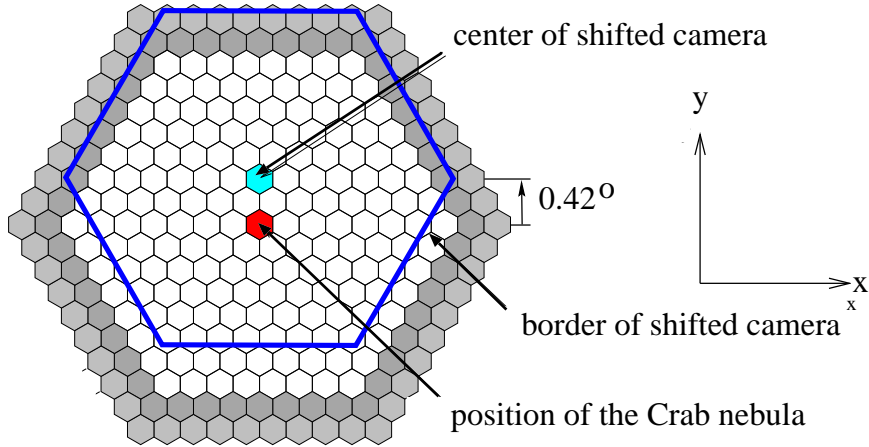}}
\end{center}
\caption
{Camera of HEGRA telescope CT3, showing the 169 pixels used in 
the initial Crab analysis (the shaded two outer rings were
excluded), and the second set of 169 pixels shifted
with respect to the center of the camera.}
\label{fig_camera}
\end{figure}

While it is unlikely that an artefact in the reconstruction procedure
generates such a narrow spike, the center of the camera is nevertheless
a preferred location and one would like to demonstrate that the 
technique works equally well for off-center sources. Lacking data
with the Crab off-center, we made use of the fact that the camera
is actually larger than the 169 pixels used in this analysis, and
selected another set of pixels shifted by about $0.42^\circ$
with respect to the center of the camera (see Fig.~\ref{fig_camera}).
The image analysis was then repeated with this ``new'' camera, with
{\em distance} cuts etc. applied relative to the new center. The result
is shown in Fig.~\ref{fig_map}(b); an excess of events is seen which
is shifted by $0.42^\circ \pm 0.01^\circ$ with respect to the new
camera center, as expected. The width of the peak is within errors
identical with the original peak. This test demonstrates that the
method is able to properly reconstruct off-center sources.

\begin{figure}[htb]
\begin{center}
\mbox{
\epsfysize10cm
\epsffile{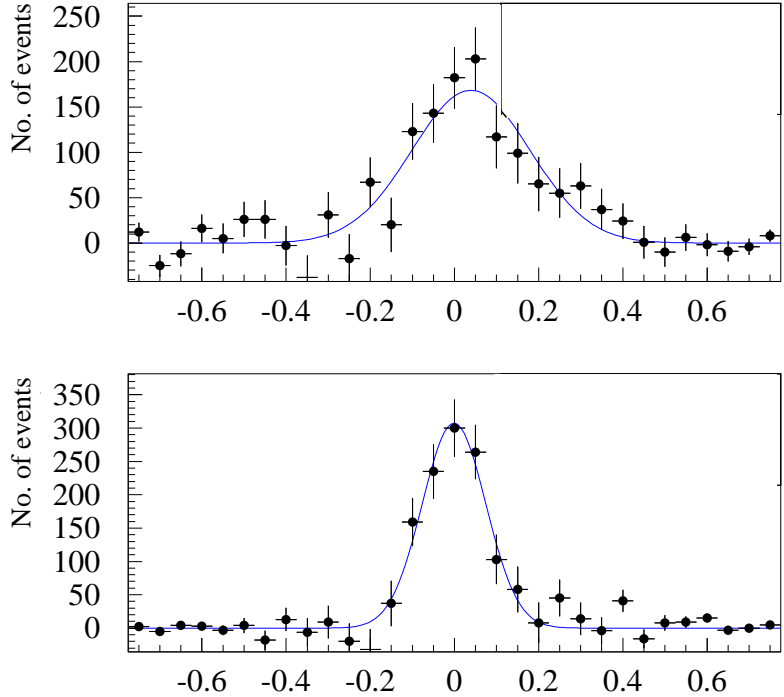}}
\end{center}
\caption
{Distribution of the Crab signal in the directions $x'$ (along
the image axis) and $y'$ (perpendicular to the image axis), after
background subtraction on the basis of off-source runs. The curves
represent gaussian fits.}
\label{fig_res}
\end{figure}

From the width of the Crab signal, the angular resolution can be 
determined. Projecting the excess on the $x$ or $y$ axis, one
finds that the distributions are within statistics described by Gaussians
with a width of $0.10^\circ \pm 0.01^\circ$ to $0.11^\circ 
\pm 0.01^\circ$, depending on the
distance cut in the pre-selection.
While the width of the excess is,
within errors, identical in $x$ and $y$, the simulations predict
that on an event-by-event basis
the reconstruction of the shower direction works better 
in the $y'$ direction, perpendicular to the image axis, than
in the $x'$ direction parallel to the image axis, where the
image shape has to be used to estimate the impact distance.
Fig.~\ref{fig_res} shows the distribution of the excess
along $x'$ and $y'$, confirming this expectation.
The corresponding angular resolutions are summarized 
in Table~\ref{tab_res}, both for the data and the simulations,
for two choices of the {\em distance} cut applied to select 
events.
Within errors, the measured angular widths agree with the 
simulation results.
\begin{table} [htb]
\begin{center}
\begin{tabular}{|l|c|c|}
\hline
 & Data & Monte Carlo \\
\hline 
\multicolumn{3}{|c|}{\em Distance $< 1.1^\circ$} \\ 
\hline
Average $(x,y)$     & $0.10^\circ \pm 0.01^\circ$ & $0.102^\circ$ \\
Longitudinal $(x')$ & $0.14^\circ \pm 0.02^\circ$ & $0.133^\circ$ \\
Transverse   $(y')$ & $0.08^\circ \pm 0.01^\circ$ & $0.086^\circ$ \\
\hline
\multicolumn{3}{|c|}{\em Distance $< 1.3^\circ$} \\ 
\hline
Average $(x,y)$     & $0.11^\circ \pm 0.01^\circ$ & $0.112^\circ$ \\
Longitudinal $(x')$ & $0.15^\circ \pm 0.02^\circ$ & $0.165^\circ$ \\
Transverse   $(y')$ & $0.08^\circ \pm 0.01^\circ$ & $0.084^\circ$ \\
\hline
\end{tabular}
\vspace{0.5cm}
\caption{Angular resultions obtained from the Crab data set,
compared to Monte-Carlo simulations. The resolutions quoted
are derived from a Gaussian fit to the distribution of 
reconstructed shower axes, projected onto the $(x,y)$ or 
$(x',y')$ directions (see Fig. 1). For the Monte Carlo
data sets, statistical errors are $0.002^\circ$ or less.}
\label{tab_res}
\end{center}
\end{table}

Due to the improved angular resolution of the fit, and the
(at least partial) resolution of head-tail ambiguities, one
should expect a corresponding increase in the significance of
signals from point sources. The Monte-Carlo studies show that,
compared to conventional analyses, the number of background 
events $N_{BG}$ after cuts is reduced by a factor 2. Since also some
of the signal events are misreconstructed and lost, the 
significance, which is governed by $N_S/\sqrt{N_{BG}}$, 
is predicted to increase by typically 30\%, rather than by a full
factor $\sqrt{2}$. In the Crab data, the background reduction
is indeed observed as predicted. The improvement in the significance
of the excess, however, amounts only to about 15\%, due to an
increased loss of signal events. Given the limited statistics 
of the Crab data set, the origin of this difference between
data and simulation could not be resolved. 

\section{Outlook}
The analysis presented in this paper emphasizes two points
\begin{itemize}
\item Modern imaging cameras provide more information
on image details, than are exploited by the second-moment
image analysis. Detailed models of the light distribution
can be used to extract additional information, and to improve
the reconstruction of the image parameters, 
in particular concerning the direction
of the primary particle. Angular resolutions can be improved
by factors 1.5 to 2.
\item While conceived for the analysis of stereoscopic
observations of air showers with multiple telescopes, 
the technique can also be applied to single-telescope data,
where it  
provides the option to generate a genuine source
map, rather than testing the hypothesis of a source
at one specific location. 
Using the relation between the shower impact parameter
and the {\em length} and {\em width} of an image, this 
technique to construct source maps can also be employed
in simplified form, based only on the second-moment
image parameters, at some expense in resolution.
\end{itemize}

\section*{Acknowledgements}

The support of the German Ministry for Research 
and Technology BMBF is acknowledged. 
We are grateful to the other members of the HEGRA collaboration,
who have participated in the development, installation, and
operation of the telescopes.
We thank the Instituto
de Astrofisica de Canarias for providing the site for 
the HEGRA IACTs, as well as excellent working conditions.

\end{document}